\magnification=\magstep1

%
\hsize = 6.50truein
\vsize = 8.50truein
\hoffset = 0.0truein
\voffset = 0.0truein
\lineskip = 2pt
\lineskiplimit = 2pt
\overfullrule = 0pt
\tolerance = 2000
\topskip = 0pt
\baselineskip = 18pt
\parindent = 0.4truein
\parskip = 0pt plus1pt
\def\medskip{\vskip6pt plus2pt minus2pt}
\def\bigskip{\vskip12pt plus4pt minus4pt}
\def\smallskip{\vskip3pt plus1pt minus1pt}
\centerline{\bf The Importance of Static Correlation in the Band}
\centerline{\bf Structure of High Temperature Superconductors}
\bigskip
\centerline{Jason K. Perry}
\centerline{\it First Principles Research, Inc.}
\centerline{\it 8391 Beverly Blvd., Suite \#171, Los Angeles, CA 90048}
\centerline{\it www.firstprinciples.com}
\bigskip
\centerline{{\it J. Phys. Chem.}, in press.}
\bigskip
\noindent{\bf Abstract:} 
\bigskip
Recently we presented a new band structure for La$_{2-x}$Sr$_x$CuO$_4$ and
other high temperature superconductors in which a second narrow band was
seen to cross the primary band at the Fermi level.  The existence of this
second Fermi level band is in complete disagreement with the commonly accepted 
LDA band structure.  Yet it provided a crucial piece of physics which led
to an explanation for superconductivity and other unusual phenomena
in these materials.  
In this work we present details as to the nature of
the failure of conventional methods in deriving the band structure of
the cuprates.  In particular, 
we use a number of chemical analogues to describe the
problem of static correlation in the band structure calculations and show
how this can be corrected with the predictable outcome of a Fermi level
band crossing.
\vfill
\eject
\noindent{\bf Introduction.}
\bigskip
Since their discovery more than a dozen years ago,$^1$ the cuprate high 
temperature superconductors have proven to be among the most unusual and
intriguing materials devised this century.  While their most obvious and
important characteristic is that they superconduct at temperatures far
in excess of the commonly accepted upper limit for conventional BCS 
superconductors, various experimental probes of their superconducting
and normal state properties have revealed anomolous behavior of a much more
general nature.  The NMR,$^2$
angle resolved photoemission (ARPES),$^3$ neutron scattering,$^4$ Josephson
tunnelling,$^5$ and IR$^6$ have all characterized these
materials as extremely exotic.  

The materials can generally be described as having two-dimensional CuO$_2$
sheets sandwiched between other metal oxide sheets which serve as charge
reservoirs.$^7$  In the case of La$_{2-x}$Sr$_x$CuO$_4$, the prototypical high
temperature superconductor, the environment around each Cu is a distorted
octahedron with the apical O's, which belong to the La/Sr/O planes, further
from the Cu center than the in-plane O's.  When the material is undoped,
$x=0$, the charge on the La is formally +3, the charge on each O is formally
-2, and the charge on the Cu is formally +2.  The Cu(II) is expected to be
in it's open-shell $d^9$ configuration, with the La and O ions closed
shell. This leads to the existence of a ``half-filled band'' 
from simple electron counting arguments.  Upon doping, substitution of
La(III) with Sr(II), Cu(III) ions are formally created as more electrons are
removed from that ``half-filled band''.  Superconductivity is observed over
the very narrow doping range of approximately $x=0.10-0.25$, with the optimal
doping ($T_c=39 K$) at $x=0.15$.$^8$

From early LDA band structure calculations it was generally concluded
that the materials were indeed very two-dimensional.$^9$  A Fermi surface 
arose from a single half-filled band composed of the anti-bonding
arrangement of the Cu $d_{x^2-y^2}$ and O $p_\sigma$ orbitals in the
signature CuO$_2$ planes, confirming simple expectations.  
However, this band structure
poses a great problem for physicists since there is virtually nothing
remarkable about it that would suggest some sort of exotic
supercondicting properties.  This has led to the development of a rather
odd attitude toward these LDA calculations.  It is clearly
agreed that they are missing some crucial physics.  Beyond that,
the overwhelming collection of unusual
data which characterizes these materials has led physicists to agree only that 
this missing physics must be deeply complicated.
Somehow in spite of these deficiencies though
the qualtitative picture of the LDA band structure
has effectively become conventional wisdom.

Yet Tahir-Kheli and Perry$^{10,11}$ recently offered a new theory of 
high temperature superconductivity which is remarkably simple and
explains substantially more than all previous theories. 
We showed that much of the confusion
about these materials stems from incorrect assumptions about their band
structure.  The LDA band structure calculations are based on the mean-field
approximation, which is known to breakdown in the limit of weakly interacting
particles.  Such is the case for the cuprates, for which it has been
well accepted that many-body effects (or dynamic correlation) are
important.
Correlation has been introduced in some models to
correct the problem, but to our knowledge this has always been done in
a limited way, applying the correction only to the three bands produced by
the Cu $d_{x^2-y^2}$ and two
O $p_\sigma$ orbitals.$^{12}$  These three band Hubbard models, which are
often reduced to one band Hubbard models, ignore the
effect that correlation has on the other bands in the material since it
is widely assumed that they are irrelevant.
Yet we have argued that this underlying assumption
that the single particle band structure is qualitatively correct is in
fact false
and such a limited approach to the incorporation of correlation
actually misses the most important consequence:
that the relative energy of the half-filled
band changes with respect to the full bands.  This is due to the improper
description of {\it static} correlation in the LDA band structure.
In our model, where the
correlation correction is applied more universally, the effect is
so dramatic that a second band appears at the Fermi level.  This is shown
in Figure 1.
This new band structure still has the approximately half-filled
2-D Cu $d_{x^2-y^2}$/O $p_\sigma$ band, but a second 
3-D Cu $d_{z^2}$/O$'$ $p_z$ band is seen at the Fermi level as
well, such that electrons are removed from both bands upon doping.
Significantly, we identified a symmetry allowed Fermi level crossing of the two 
bands which we showed was the crucial element in understanding the 
physics of these materials.  This band crossing allows for the formation 
of a new type of interband Cooper pair, representing a simple twist on
the conventional BCS theory of superconductivty.  Moreover, the wealth
of experimental data which demonstrates more general anomolous behavior
can easily be explained by this unusual band structure, and in a number of
cases has already been quantitatively reproduced.$^{10,11,13,14}$

In this work, we present arguments as to the nature of the correlation problem 
in conventional
LDA calculations and why correcting this problem intuitively leads to the
new band structure.  We develop these arguments from a chemist's perspective
using a number of familiar molecular systems to illustrate various aspects
of the correlation problem.  In particular, the chemistry of H$_2$, benzene,
and the Cu ion dimer will be discussed, leading up to a discussion of
the band structure for La$_{2-x}$Sr$_x$CuO$_4$.
\bigskip
\noindent{\bf The Problem with H$_2$ Dissociation}
\bigskip
To understand the basic problem with the LDA band structure calculations of
the cuprates, it is only necessary to consider the fundamental problem of
dissociation consistency in single configuration based methods.$^{15}$  
In Figure
2 we show the dissociation curves for H$_2$ as calculated at the Hartree-Fock
(HF) and the B3LYP$^{16}$ 
density functional (DFT) levels using both a restricted spin and
symmetry approach and an unrestricted spin and symmetry approach.  For
both methods (and the case is the same for other DFT functionals) the
restricted approach leads to dissociation to an excited state description
of two H atoms.  The unrestricted approach leads to proper dissociation.
This behavior is well understood and represents the primary motivation
behind the development of methods such as Generalized Valence 
Bond (GVB).$^{17}$

The problem with the restricted approach is that two electrons are forced
to occupy the same orbital (the $\sigma_g$ orbital in the case of H$_2$).  
This is a fine approximation near the equilibrium
bond length, and indeed both the restricted and unrestricted approaches
lead to the same state in this region.  However, upon dissociation,
forcing two electrons to occupy the same orbital is clearly not appropriate
since the local representation of this can be seen to be 50\% covalent
(the correct dissociation limit) and 50\% ionic (an excited state).
Explicitly, that is

$$\eqalign{\Psi_g =& (\sigma_g)^2 \cr
=& {1 \over 2}(1s(H1)+1s(H2))^2 \cr
=& {1 \over 2}((1s(H1))^2 + (1s(H2))^2 + (1s(H1))^1(1s(H2))^1 +
(1s(H2))^1(1s(H1))^1) \cr
=& {1 \over {\sqrt{2}}}(\Psi(IONIC) + \Psi(COVALENT))}$$

\noindent
For the HF wavefunction, the energy of this state is

$$E_g(r=\infty) = 2E_{1s} + {1 \over 2}J_{1s,1s}$$

\noindent
where $E_{1s}$ is the ground state energy of an H atom and $J_{1s,1s}$ is the
self-Coulomb energy associated with the H $1s$ orbital.  The situation is
similar for DFT where the exchange and correlation
functionals will cancel some but not all of the self-Coulomb term.  As
a result the error for HF (7.1 eV) is seen to be larger than that for
B3LYP (2.8 eV), but the error for B3LYP and other DFT functionals is
nevertheless non-zero.

The unrestricted approach overcomes the problem of the self-Coulomb
energy by breaking spin and symmetry and localizing the alpha spin
electron on one H atom and the beta spin electron on the other.  As a
result, there is dissociation to the proper $\Psi(COVALENT)$ limit.  
Alternatively, a
method which introduces static correlation, such as GVB 
(or more generally CASSCF), overcomes this problem
without breaking spin by describing the bond with two configurations
as,

$$\Psi_{GVB} = c_1(\sigma_g)^2 - c_2(\sigma_u)^2$$

\noindent
where $c_1^2 + c_2^2 = 1$.  The energy upon dissociation is,

$$E_{GVB}(r=\infty) = c_1^2E_g(r=\infty) + c_2^2E_u(r=\infty) - c_1c_2J_{1s,1s}$$

\noindent
Clearly since $E_g = E_u$ upon dissociation, the optimal set of coefficients
is $c_1 = c_2 = {1 \over{\sqrt{2}}}$.  Hence the GVB wavefunction
dissociates properly to

$$E_{GVB}(r=\infty) = 2E(1s)$$

While this is all very familiar, the point is that it is pertinant to
the electronic structure of high temperature superconductors.  In these
materials the Cu(II) $d^9$ spins of the half-filled band
are separated by 3.8 \AA.  At this separation,
a breakdown in the mean-field approximation is expected resulting in
a substantial overestimate of the self-Coulomb term.
Recognition of such has been the motivation behind
calculations in which the La$_2$CuO$_4$ unit cell has been doubled to
allow for spin polarization.$^{18,19}$  
In these calculations alpha and beta spins
localize to alternating sites in the undoped material, thus 
removing the self-Coulomb
term associated with the half-filled band much like the unrestricted
spin and symmetry calculations remove the self-Coulomb term from dissociated
H$_2$.  The work of Svane$^{18}$ is
particularly important in this regard since it also accounts for the
fact that the
self-Coulomb term and the self-exchange and correlation terms do not
completely cancel.  As a solution, 
he applies a self-interaction correction (SIC) to
those orbitals that can be well localized.  While in context this is
correct, and to some extent his calculations are in agreement with ours,
as we show next, correlation of delocalized orbitals is important, too.
\bigskip
\noindent{\bf Static Correlation in Benzene}
\bigskip
A more complicated example of static correlation is the case of aromatic
benzene.  At the HF level, there are three orbitals having the symmetries
$A_{2u}$ and $E_{1g}$ under the $D_{6h}$ point group which represent the
delocalized form of the three benzene $\pi$ orbitals.  Yet the six
atomic p$_\pi$ orbitals which form these three molecular
orbitals have only a moderate overlap with each other.  This leads to an
overestimate of the self-Coulomb term associated with the bonds which
requires correlation of the type just described.
The easiest way to introduce such correlation
is through the GVB approach in which symmetry is broken and
the three delocalized HF $\pi$ orbitals 
are localized to three $\pi$ bonds corresponding to one of the
two resonating
Kekul\'e structures.  Similarly, the three corresponding $\pi$ antibonding
orbitals are localized and the GVB wavefunction becomes

$$\Psi_{GVB} = (c_1(\pi_g(1))^2 - c_2(\pi_u(1))^2)(c_1(\pi_g(2))^2
- c_2(\pi_u(2))^2)(c_1(\pi_g(3))^2 - c_2(\pi_u(3))^2)$$

The energy of the GVB wavefunction is 1.12 eV lower than that of the HF
wavefunction using a 6-311G** basis set.$^{20}$  This represents a lowering of
0.37 eV per bond, which can be directly related to a reduction in the
self-Coulomb term associated with each bond.

Additional correlation to account for spin-polarization of the bonds
can be introduced through the RCI wavefunction which adds the
single excitation configuration $c_3(\pi_g)^1(\pi_u)^1$ for
each bond in the above equation for $\Psi_{GVB}$ while also relaxing some
inherent constraints on the GVB coefficients.  This correlation
effectively allows alpha and beta spins to separate and lowers the
total energy by another 0.30 eV.

Resonance can then be included by allowing all excitations between the
bonds ({\it i.e.} all excitations of the six electrons within the space
of the six GVB orbitals).  This
GVBCI wavefunction lowers the total energy by another 0.53 eV.  Significantly,
this GVBCI wavefunction is also strictly equivalent to the commonly used CASSCF 
wavefunction.  The two are related by a simple transformation from the
localized space (GVBCI) to the delocalized space (CASSCF).  The very
existence of this
transformation implies that the correlation which exists in the GVBCI
also exists in the CASSCF.  Since it is clear that the most important
correlation in the GVBCI is that
which reduces the self-Coulomb energy of the $\pi$ bonds, the same
must be true of the CASSCF, although it is much less transparent.  
In other words, the correlation which reduces the self-Coulomb energy
is independent
of whether the orbitals are localized or delocalized.

The presence of this same type of correlation in systems that are
delocalized is often overlooked.  In the case of the superconductors,
methods that depend on the localization of orbitals in order to reduce the
self-Coulomb energy$^{18}$ are in fact biased toward such well localized states
since they miss the fact that the energy can be similarly lowered by
application of such correlation to states that cannot be well
localized.  This is not
to say that undoped La$_2$CuO$_4$ does not in fact have well localized spins,
since the undoped 
material is clearly an antiferromagnet.  But upon doping, when
orbitals can no longer be easily localized, this type of correlation should
not be expected to just disappear.  By our argument here, reduction of the
self-Coulomb energy should be considered for both localized and delocalized
orbitals in
evaluating the band structure.  The consequences of this are addressed
in the next section.
\bigskip
\noindent{\bf The Problem with Separated Cu Ions}
\bigskip
The ground state of Cu(I) is known to be $^1S$ $d^{10}$, the ground state
of Cu(II) is known to be $^2D$ $d^9$, and the ground state of Cu(III) is
known to be $^3F$ $d^8$.$^{21}$  While it is the case that there is only one
possible $d^{10}$ configuration for Cu(I), and the five possible
$d^9$ configurations for Cu(II) are degenerate, for Cu(III) the ten
different possible triplet $d^8$ configurations lead to different mixtures of
the $^3F$ and higher energy $^3P$ states.  Only the two configurations
in which one hole is in the $d_{\sigma}$ orbital and the other is
in a $d_{\delta}$ orbital lead to a pure $^3F$ state in a single
reference description.  

Using a triple$-\zeta$ contraction of Hay and Wadt's
ECP basis set,$^{22}$ 
we calculate a second ionization potential (the difference
between Cu(I) and Cu(II)) to be 17.54 eV at the HF level and 20.65 eV
at the B3LYP level in comparison to the experimental value of 20.29 eV.
Similarly, we calculate a third ionization potential (the difference between
Cu(II) and Cu(III)) to be 34.32 eV at the HF level and 37.06 eV at
the B3LYP level in comparison to the experimental value of 36.83 eV.  Clearly,
B3LYP is a suitable method for studying the Cu ions.

Yet we find that when two Cu ions are low spin coupled and separated by a long 
distance, these methods have difficulty.  As with H$_2$, an unrestricted
spin and symmetry approach will properly describe the two ions, but
attempting to use a restricted spin and symmetry approach fails.  The
nature of this failure is quite revealing however in how it relates to
the band structure of the high temperature superconductors.

Results of calculations on various Cu ion dimers are given in Table I.
As can be seen, the energy of the Cu(I)$+$Cu(I) dimer where both ions
are $d^{10}$ is correct.  The energy of the Cu(I)$+$Cu(II) dimer where
each ion is an average of $d^9$ and $d^{10}$ is also correct.  However,
the energy of the singlet state of Cu(II)$+$Cu(II) is high by 14.42 eV
at the HF level and high by 4.30 eV at the B3LYP level.  This
state has the following orbital occupations

$$Cu(II)+Cu(II) = 
(xy_g)^2(xy_u)^2(xz_g)^2(xz_u)^2(yz_g)^2(yz_u)^2(z^2_g)^2(z^2_u)^2(x^2-y^2_g)^2(x^2-y^2_u)^0$$

\noindent
As shown for
H$_2$ and benzene, the error in the Cu(II)$+$Cu(II) energy
can be unambiguously attributed to the
lack of static correlation in the half-filled $d_{x^2-y^2}$ 
pair of orbitals which
leads to this copper dimer being described as 50\% Cu(II)$+$Cu(II)
and 50\% Cu(I)$+$Cu(III).  This state can be correctly described by the GVB
or CASSCF method or by breaking symmetry and spin in an unrestricted
approach.  Alternatively, changing the spin to triplet and singly occupying
each of the two $d_{x^2-y^2}$ orbitals will lead to the correct ground
state.

This Cu(II)$+$Cu(II) model by itself offers a good argument for what might be
wrong with conventional LDA band structure calculations of the cuprate
superconductors.  Doubling the unit cell to allow breaking of symmetry and
spin with localization of the alpha and beta spins on alternating copper
sites may be one logical solution for understanding the undoped material.
Alternatively, introducing more rigorous correlation with a Hubbard model
of the isolated Cu $d_{x^2-y^2}$ / O p$_{\sigma}$ band may be another
logical solution.  However when our model is taken one step further to
consider Cu(II)$+$Cu(III) the most important aspect of the lack of
static correlation in the half-filled band can be seen, and this point has
received little attention until now.

When one more electron is removed from the $d_{x^2-y^2}$ pair of orbitals
to form Cu(II)$+$Cu(III), the doublet state is again described {\it correctly}
even though it corresponds to an excited state configuration of Cu(III).  The
state is actually $^2D$ Cu(II) $+$ $^1G$ Cu(III), where the $^1G$ $d^8$
configuration of Cu(III) corresponds to having the $d_{x^2-y^2}$ orbital
empty.  We calculate the
$^3F$ $\rightarrow$ $^1G$ excitation energy to be 4.29 eV at the HF level and
3.86 eV at the B3LYP level.  However, when an electron is instead removed
from the $d_{z^2}$ pair of orbitals, which should lead to a ground
state description of $^2D$ Cu(II) $+$ $^3F$ Cu(III), 
the doublet coupling of the two ions is {\it too high}
in energy by 15.67 eV at the HF level and 5.03 eV at the B3LYP level.
Even correcting the improper exchange interaction between the $d_{z^2}$
and $d_{x^2-y^2}$ electrons in this configuration, the HF energy is
still 14.95 eV too high, and the B3LYP energy is still 4.48 eV too high.

The difference between these two states of Cu(II)$+$Cu(III) can be
understood in that removing an electron from the $d_{x^2-y^2}$ orbitals
removes the problem with static correlation whereas removing an electron
from the $d_{z^2}$ orbitals does not.  In the former case, there is
only one electron remaining in the $d_{x^2-y^2}$ orbitals and it is shared 
equally between the two ions.  In the latter case, there are still two
electrons in the $d_{x^2-y^2}$ orbitals and without proper correlation
the self-Coulomb energy will remain too high.  In the end, this means
that in starting with a half-filled set of $d_{x^2-y^2}$ orbitals in
Cu(II)$+$Cu(II) there is an improper bias of 14.42 eV at the HF level
and 4.30 eV at the B3LYP level {\it toward} removing an additional electron
from $d_{x^2-y^2}$.  However there is actually a bias of 0.53 eV at the
HF level and 0.18 eV at the B3LYP level {\it against} removing an
electron from $d_{z^2}$.  In other words, the lack of correlation in
the $d_{x^2-y^2}$ orbitals raises the energy of those particular orbitals with
respect to all the other orbitals.

The three models discussed here, (H$_2$, benzene, and the Cu ion dimer),
suggest that static correlation needs to be considered in the band structure
of the cuprate superconductors, that it needs to be applied to all orbitals
regardless of whether or not they can be well localized, and that the
primary result will surely be to lower the energy of the entire half-filled
band with respect to the other filled bands.
\bigskip
\noindent{\bf The Importance of Static Correlation in the Band Structure of
High Temperature Superconductors}
\bigskip
We have chosen to study the band structure of optimally doped 
La$_{2-x}$Sr$_x$CuO$_4$ with a Hubbard model which uses parameters
derived from DFT calculations on a CuO$_6$ cluster.  The details of
the cluster calculations and the procedure for extracting the Hubbard
parameters are given explicitly in Perry and Tahir-Kheli.$^{11}$  
All parameters
necessary to describe the Cu $d_{x^2-y^2}$ / O $p_\sigma$ and Cu $d_{z^2}$ /
O$'$ $p_z$ bands were derived.  These parameters include orbital energies,
Coulomb and exchange energies, and orbital couplings.  Our original
set of parameters, which were published in that work, came from BLYP/6-31+G*
calculations (using an ECP on the Cu).  We have since derived parameters
from B3LYP/6-311+G* calculations and found the resulting 2-D
band structure
(detailed below) to be qualitatively the same as that obtained with the earlier
parameter set.  However, we have also included a 3-D
coupling in this new band structure and as a result we can now calculate
such experimental observables as the NMR Cu and O spin relaxation rates,$^{13}$
the ARPES Fermi surface, the neutron scattering, and the mid-IR 
absorption$^{14}$
with near quantitative accuracy, something that has not been done with
any other band structure.

The validity of the general approach can be tested by calculating the Hubbard
model band structure within the mean-field approximation.  The calculation
must be done iteratively until self-consistency is achieved because the
orbital energies depend on the Coulomb and exchange field which depends on
the orbital occupations which depend on the orbital energies.  The first
step is to calculate the orbital energies as a function of the orbital
occupations.  Under the mean-field approximation, this is

$$E_i = E_i^0 - \sum_j (2-N_j)(J_{i,j} - {1 \over 2} K_{i,j})$$

\noindent
where $E_i^0$ are the calculated orbital energies when all valence bands are
full (formally La(III), Sr(II), Cu(I), and O(-II)), $N_j$ are the atomic
orbital occupations, $J_{i,j}$ are the Coulomb terms between orbitals, and
$K_{i,j}$ are the exchange terms.  Details of how the long range Coulomb
field is handled are given in the cited reference.$^{11}$  Once the orbital
energies are determined, a Hubbard matrix is constructed at every {\bf k}
vector on a grid covering the first Brillouin zone, the eigenvectors and
eigenvalues of each matrix are determined corresponding to the orbitals and
orbital energies at each {\bf k} point, the Fermi level is adjusted
such that
the correct number of orbitals are occupied for the particular doping level,
the atomic orbital occupations are then determined, and the process is
repeated.  It should be noted that in our model $J_{i,i} = K_{i,i}$ such
that when an orbital is half-occupied its energy is $E_i = E_i^0 - 
{1 \over 2} J_{i,i}$.

As shown in Figure 1a, using the mean-field approximation to determine
orbital energies as above and constraining the model to a 2-D
description of the material leads to a band structure which is nearly
quantitatively identical to those published using conventional LDA
band structure techniques.$^9$ A single Cu $d_{x^2-y^2}$ / O $p_{\sigma}$
band which is widely dispersing is seen to cross the Fermi level.  A
second Cu $d_{z^2}$ / O$'$ $p_z$ band is seen to be several eV lower in
energy.  This good agreement effectively validates the procedure.

It is interesting to note however 
that the bottom of the $d_{z^2}$ band is several
eV below the bottom of the $d_{x^2-y^2}$ band even though at {\bf k}=(0,0)
the $d_{x^2-y^2}$ orbital represents a non-bonding combination of the Cu
orbitals, having no O $p_{\sigma}$ character at all, 
while the $d_{z^2}$ orbital
has significant anti-bonding O$'$ $p_z$ character.  Ligand field theory would
suggest that the $d_{z^2}$ band 
should be higher in energy than the $d_{x^2-y^2}$
band at this {\bf k} point unless the $d_{z^2}$ atomic orbital is itself
significantly more stable than the $d_{x^2-y^2}$ atomic orbital.  This is
indeed the case, but it
cannot be explained by differences in the intrinsic $E_i^0$ 
atomic orbital energies for $d_{z^2}$ and $d_{x^2-y^2}$ 
since this difference is only 0.13 eV.
The stabilization of the $d_{z^2}$ band with respect to the $d_{x^2-y^2}$
band is seen only upon removal of electrons from the $d_{x^2-y^2}$ band.
This is counterintuitive and exactly the opposite behavior should be
expected from such basic principles as Hund's rule.  
It is a direct result though of the improper accounting
of the self-Coulomb energy in the mean-field approximation for this
strongly correlated system.  This behavior is completely analogous to that
seen for the Cu ion dimer discussed above.  Thus we expect that correlation
that would reduce the self-Coulomb term of partially occupied orbitals would
lower the energy of the Cu $d_{x^2-y^2}$ orbital
with respect to the Cu $d_{z^2}$ orbital.

Introducing static correlation to the band structure in a rigorous way
is an extremely difficult problem.  However, the effect of this
correlation on the self-Coulomb term in the mean-field equation can
easily be approximated.  This is best seen by considering Figure 3 and
thinking about what the self-Coulomb energy should be when a particular
atomic orbital is half-filled.

Figure 3 depicts a localized description of the Cu $d_{x^2-y^2}$ / 
O $p_{\sigma}$ band.  Such localization can be exact only when the band is
half-filled.  The localization can still be approximately correct with
the addition or removal of electrons if the ensuing delocalized states
are viewed as arising from the resonance of localized states.
Figure 3a shows the mean-field spin coupling in the CuO$_2$ plane
while Figure 3b shows an antiferromagnetic spin coupling which is relevant when
the material is undoped.  Upon doping, this antiferromagnetic order is
destroyed and a correlated paramagnetic spin-coupling such as that depicted in
Figures 3c and 3d is expected.

In the mean-field picture, when
the Cu $d_{x^2-y^2}$ orbital is half occupied, the local spin is 50\%
alpha and 50\% beta leading to a self-Coulomb term which is ${1 \over 2} J$.
However, in both the antiferromagnetic and correlated paramagnetic
pictures when the Cu $d_{x^2-y^2}$
orbital is half occupied,  a resonance exists between states that have
a local spin in that orbital that is purely alpha or purely beta.  This
picture is fundamentally different from that of the mean-field approximation
and leads to a self-Coulomb term which is $0 J$.  From the arguments
used to make the connection between the GVBCI and CASSCF descriptions of
benzene, the same can be said of the Cu $d_{z^2}$ and O$'$ $p_z$ orbitals 
even though localization of these orbitals is not as straightforward.
That is, delocalized states must be viewed as arising from the resonance of 
very low symmetry localized
states.  So for the Cu $d_{x^2-y^2}$ and $d_{z^2}$ orbitals and the O$'$ $p_z$
orbital, the correlation corrected mean-field equation becomes

$$E_i = E_i^0 - (2 - N_i)J_{ii} - \sum_{j{\neq}i} (2-N_j)(J_{ij} -
{1 \over 2} K_{ij}),\ \ \ \ \ N_i > 1$$

$$E_i = E_i^0 - J_{ii} - \sum_{j{\neq}i} (2-N_j)(J_{ij} - {1 \over 2} K_{ij}),
\ \ \ \ \ \ \ \ \ \ \ \ \ \ \ \ N_i {\leq} 1$$

\noindent
Upon examination, it can easily be seen that if an orbital is half-occupied
or less, the full self-Coulomb term will be removed from $E_i^0$.

The situation is a little less clearcut for the O $p_{\sigma}$ orbitals.
In the antiferromagnetic picture of Figure 3b, alpha or beta spin is
localized to alternating Cu sites, but as a result each O site is then
50\% alpha and 50\% beta.
Thus, the self-Coulomb term is expected to be ${1 \over 2} J$
for the half-occupied orbital as it is under the mean-field approximation.
In the correlated paramagnetic picture of Figure 3c and 3d, for the
one O atom that lies between two spin paired Cu atoms, the self-Coulomb
term also turns out to be ${1 \over 2} J$.  However, for the three other
O atoms surrounding any particular Cu site, the self-Coulomb term is
expected to be ${1 \over 4} J$.  This is because the uncorrelated spins
between the two Cu atoms lead to spin on the O which is 25\% pure alpha,
25\% pure beta, and 50\% half alpha/half beta.  The latter term leads to
the ${1 \over 4} J$ Coulomb repulsion.  On average then, when the O
$p_{\sigma}$ orbital is half-occupied, the self-Coulomb term is
${3 \over 4}\times{1 \over 4} J + {1 \over 4}\times{1 \over 2} J = {5 \over
16} J$.  The correlation corrected mean-field equation for this orbital
then becomes

$$E_i = E_i^0 - {11 \over 16}(2 - N_i)J_{ii} - \sum_{j{\neq}i}
(2-N_j)(J_{ij} - {1 \over 2} K_{ij}),\ \ \ \ \ \ \ \ \ \ N_i > 1$$

$$E_i = E_i^0 - ({5 \over 16}(2 - N_i) + {3 \over 8})J_{ii} - 
\sum_{j{\neq}i} (2-N_j)(J_{ij} - {1 \over 2} K_{ij}),
\ \ \ \ \ N_i {\leq} 1$$

\noindent
This latter set of equations is clearly approximate and may vary substantially
from that obtained from the exact wavefunction, which is of course unknown.
So we should note that we have generated band structures with a variety of
values for the extent of the self-Coulomb term removed from the O $p_{\sigma}$
$E_i^0$ atomic orbital energies to test the importance of this term.
For values ranging from ${1 \over 2} J$ removed at half-occupancy to 
a full $J$ removed, 
no qualitative difference in the band structure was observed.
We thus feel that the choice of ${11 \over 16} J$ removed from the orbital
energy for O $p_{\sigma}$ at half-occupany is reasonable.

The results of including this static correlation in the Hubbard model can
be seen in Figure 1b.  Here we present the two-dimensional band structure
obtained with the newer B3LYP/6-311+G* parameters.  As occurs with the older
BLYP/6-31+G* band structure, the Cu $d_{x^2-y^2}$ / O $p_{\sigma}$ band
is seen to be stabilized with respect to the Cu $d_{z^2}$ / O$'$ $p_z$ band.
The change is so dramatic that the second band is seen now to lie just below
the Fermi level at optimal doping, a rather robust effect.  As we pointed
out in our first published work on this subject, a symmetry allowed crossing
of the two bands is observed very near the Fermi level.$^{11}$  For this 
newer set of
parameters, it is just 0.024 eV below the Fermi level.  The existence of
such a crossing provides the unique opportunity for a new type of Cooper
pair to form.  In conventional BCS superconductors, pairs of electrons
near the Fermi level form an attractive coupling when one of the electrons is
in state {\bf k} and the other is in state {\bf -k}.  With the existence of
a Fermi level band crossing, such an attractive coupling can be formed
between electrons in states {\bf k} and {\bf -k} where each of the electrons
belongs to a {\it different} band.  This new and simple twist on the
conventional theory immediately provides an explanation for the $d$-wave
gap observed in the Josephson tunneling$^{5,10}$ and ARPES.$^{3,14a}$

While our early work resorted to empirical modifications to the Hubbard
model to a achieve a band crossing at exactly the Fermi level, recently
we found that the introduction of a small 3-D coupling
on the order of 0.05-0.15 eV between O$'$ $p_z$ orbitals of neighboring
planes was enough to produce a Fermi level band crossing.$^{14c}$  This is shown
in Figure 4.  The crossing occurs in a limited area of the 3-D
Brillouin zone, but this is all that is necessary for the formation of
interband Cooper pairs.  
We should mention that several researchers have previously 
noted $z^2$ character near the Fermi level in spin-polarized band structure
calculations on undoped La$_2$CuO$_4$,$^{18,19}$ 
so this new band structure should
not come as a complete surprise even though it is radically different from
the band structure that has gained common acceptance.  
To our knowledge though, no one has ever noted
the band crossing before, and it is this that leads directly to the
unusual physics of high temperature superconductivity.
\bigskip
\noindent{\bf Conclusions}
\bigskip
We have shown that the conventional LDA band structure calculations for 
La$_{2-x}$Sr$_x$CuO$_4$ and other high temperature superconductors have
failed due to an underestimation of the static correlation.  This same
failure affects molecular systems such as H$_2$, benzene, and the Cu ion
dimer in a well understood way.  We have corrected the problem within the
framework of a Hubbard model by altering the accounting associated with
the self-Coulomb term.  The result was a radically different band structure
in which a second Cu $d_{z^2}$ / O$'$ $p_z$ band was seen to cross the
primary Cu $d_{x^2-y^2}$ / O $p_{\sigma}$ band at the Fermi level.  The
observation of this band crossing led to a new interband pairing theory
for the mechanism of superconductivity in these materials.

Finally, we must stress that not only does the new band structure
and interband pairing theory explain the origin of $d$-wave superconductivity
in these materials, it explains the origin of the high T$_c$ as resulting
from unusual behavior in the dielectric constant stemming from the band
crossing.$^{14b}$  It also quantitatively explains the anomolous behavior of the
NMR Cu and O spin relaxation rates as simply the result of rapidly changing
orbital character near the Fermi level.$^{13}$  It explains the ARPES pseudogap
as originating from the very narrowly dispersing Cu $d_{z^2}$ band.$^{14a}$  It
further explains the incomensurate peaks of the neutron scattering and
the mid-IR absorption.$^{14d}$  None of the physics associated with
understanding these experiments is particularly difficult when this new
band structure is used.  In contrast, the physics that has been proposed
by various sources 
in reference to the conventional band structure to explain any one of the
above mentioned experiments has always been deeply complex and limited
in its predictive capability.  We suggest that nature usually prefers
the simpler solution.
\bigskip
\noindent{\bf Acknowledgment}
\bigskip
We would like to acknowledge the substantial contributions of 
Dr. Jamil Tahir-Kheli
to this work.  We would also like to acknowledge many useful discussions
with Dr. Jean-Marc Langlois.  We especially thank Prof. William A. Goddard III
for his continuing guidance.
\vfill
\eject
\noindent{\bf References.}

\item{$^1$}J.G. Bednorz and K.A. M\"uller, Z. Phys. B {\bf 64}, 189 (1986).

\item{$^2$}R.E. Walstedt, B.S. Shastry, and S.-W. Cheong, {\it Phys. Rev.
Lett.} {\bf 72}, 3610 (1994).

\item{$^3$}M.R. Norman, H. Ding, M. Randeria, J.C. Campuzano, T. Yokoya,
T. Takeuchi, T. Takahashi, T. Mochiku, K. Kadowaki, P. Guptasarma, and
D.G. Hinks, {\it Nature} {\bf 392}, 157 (1998).

\item{$^4$}P. Dai, H.A. Mook, and F. Dogan, {\it Phys. Rev. Lett.} {\bf 80},
1738 (1998).

\item{$^5$}C.C. Tsuei, J.R. Kirtley, C.C. Chi, L.S. Yu-Jahnes, A. Gupta, T.
Shaw, J.Z. Sun, and M.B. Ketchen, {\it Phys. Rev. Lett.} {\bf 73}, 593 (1994).

\item{$^6$}D.B. Tanner and T. Timusk, in {\it Physical Properties of High
Temperature Superconductors III}, ed. D.M. Ginsberg (World Scientific, New
Jersey; 1990), 363.

\item{$^7$} R.M. Hazen, in {\it Physical Properties of High Temperature
Superconductors II}, ed. D.M. Ginsberg (World Scientific, New Jersey; 1990),
121.

\item{$^8$}H. Takagi, R.J. Cava, M. Marezio, B. Batlogg, J.J. Krajewski,
W.F.Peck, Jr., P. Bordet, D.E. Cox, Phys. Rev. Lett. {\bf 68}, 3777 (1992).

\item{$^9$}J. Yu, A.J. Freeman, and J.H. Xu, Phys. Rev. Lett. {\bf 58},
1035 (1987);
L.F. Mattheiss, Phys. Rev. Lett. {\bf 58}, 1028 (1987);
W.E. Pickett, Rev. Mod. Phys. {\bf 61}, 433 (1989), and
references therein.

\item{$^{10}$}J. Tahir-Kheli, {\it Phys. Rev. B} {\bf 58}, 12307 (1998).

\item{$^{11}$}J.K. Perry and J. Tahir-Kheli, {\it Phys. Rev. B} {\bf 58},
12323 (1998).

\item{$^{12}$}M.S. Hybertsen, E.B. Stechel, W.M.C. Foulkes, and M. Schl\"uter,
{\it Phys. Rev.  B} {\bf 45}, 10032 (1992).

\item{$^{13}$}J. Tahir-Kheli, J. Phys. Chem., in press.

\item{$^{14}$}a) J.K. Perry and J. Tahir-Kheli, submitted to Phys. Rev. Lett.; 
b) J. Tahir-Kheli, submitted to Phys. Rev. Lett.;
c) J.K. Perry and J. Tahir-Kheli, submitted to Phys. Rev. Lett.;
d) J. Tahir-Kheli, to be published.  See www.firstprinciples.com.

\item{$^{15}$}The reader is referred to any number of standard texts such
as I.N. Levine, {\it Quantum Chemistry, 4th ed} (Prentice Hall, Englewood
Cliffs, New Jersey; 1991); A. Szabo and N.S. Ostlund, {\it Modern Quantum
Chemistry: Introduction to Advanced Electronic Structure Theory, 1st ed. rev.}
(McGraw-Hill, New York; 1989).

\item{$^{16}$}A.D. Becke, {\it J. Chem. Phys.} {\bf 98}, 5648 (1993).

\item{$^{17}$}F.W. Bobrowicz and W.A. Goddard III, in {\it Modern Theoretical
Chemistry: Methods of Electronic Structure Theory}, edited by H.F. Schaefer III
(Plenum, New York; 1977).

\item{$^{18}$}A. Svane, Phys. Rev. Lett. {\bf 68}, 1900 (1992).

\item{$^{19}$}K. Shiraishi, A. Oshiyama, N. Shima, T. Nakayama, and H.
Kamimura, Solid State Comm. {\bf 66}, 629 (1988).

\item{$^{20}$}R. Krishnan, J.S. Binkley, R. Seeger, and J.A. Pople, {\it
J. Chem. Phys.} {\bf 72}, 650 (1980).

\item{$^{21}$}C.E. Moore, {\it Atomic Energy Levels}, NSRDS-NBS 35 (reprint of
NBS circular 467) (U.S. Government Printing Office, Washington, D.C., 1971).

\item{$^{22}$}P.J. Hay and W.R. Wadt, {\it J. Chem. Phys.} {\bf 82}, 299
(1985).
\vfill
\eject
\noindent{\bf Table I.} Calculated energetics for the Cu ion dimer (in eV).
HF(calc) and B3LYP(calc) are computed under a spin and symmetry
restricted formalism.  HF(exact) and B3LYP(exact) represent the correct
values for two non-interacting ions.
\vskip 0.5truein
\halign{\noindent#\hfill &\quad \hfill#\hfill &\quad \hfill#\hfill
&\quad \hfill#\hfill &\quad \hfill#\hfill \cr
\noalign{\bigskip\hrule\smallskip}
\noalign{\hrule\medskip}
Dimer & HF(calc) & HF(exact) & B3LYP(calc) & B3LYP(exact) \cr
\noalign{\medskip\hrule\medskip}
$Cu(I/^1S)+Cu(I/^1S)$ & 0.00 & 0.00 & 0.00 & 0.00 \cr
$Cu(I^1S)+Cu(II/^2D)$ & 17.54 & 17.54 & 20.65 & 20.65 \cr
$Cu(II/^2D)+Cu(II/^2D)$ & 49.50 & 35.08 & 45.60 & 41.30 \cr
$Cu(II/^2D)+Cu(III/^1G)$ & 56.15 & 56.15 & 61.57 & 61.57 \cr
$Cu(II/^2D)+Cu(III/^3F)$ & 67.53 & 51.86 & 62.74 & 57.71 \cr
\noalign{\medskip\hrule\smallskip}
\noalign{\hrule\bigskip}}
\vfill
\eject
\noindent{\bf Figure Captions}
\bigskip
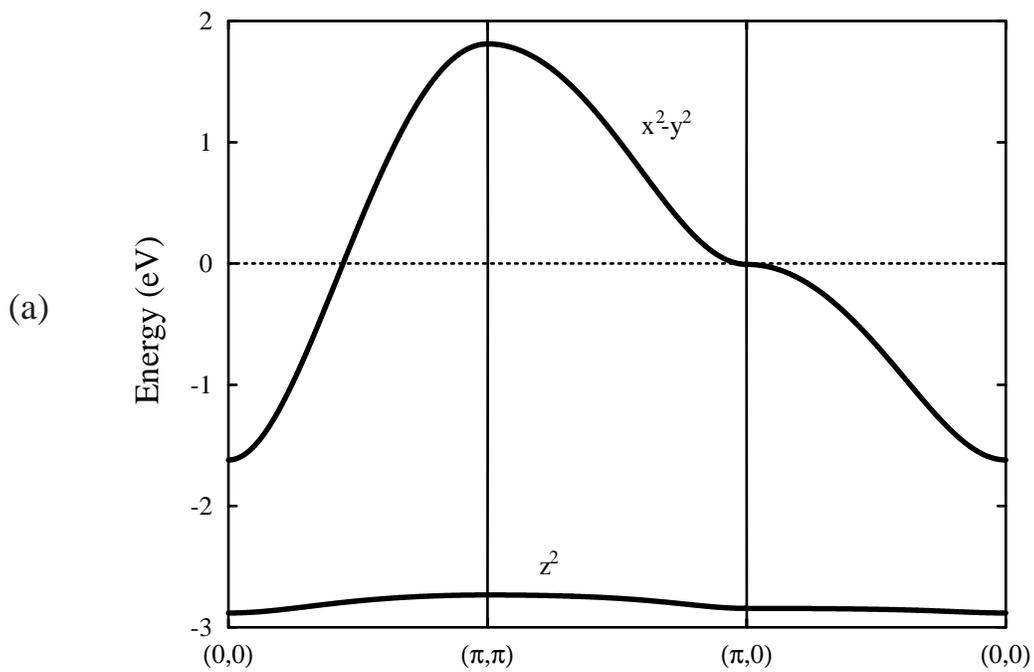
\noindent{\bf Figure 1.} {\bf a}  Calculated 2-D band structure for optimally
doped La$_{1.85}$Sr$_{0.15}$CuO$_4$ using our Hubbard model and retaining
the mean-field approximation.  {\bf b}  Calculated 2-D band structure using
our Hubbard model and including static correlation.  The two bands are
seen to cross along the $(0,0)-(\pi,\pi)$ direction very near the Fermi
level.  Note other bands are not shown for clarity.
\bigskip
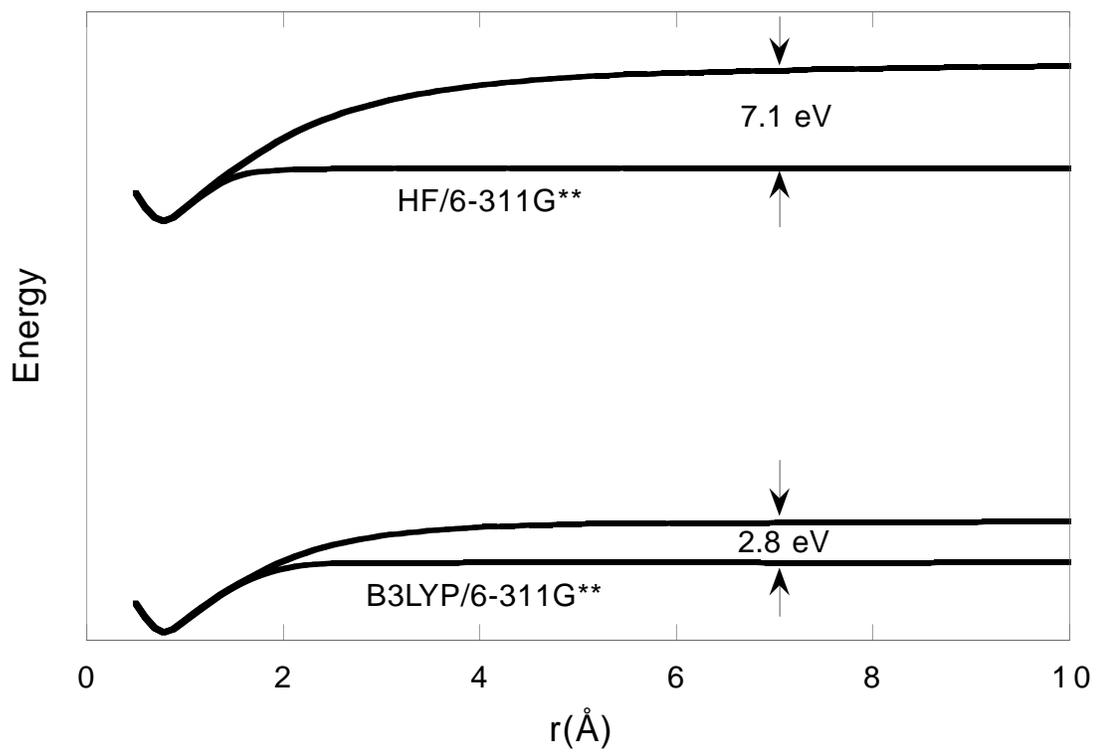
\noindent{\bf Figure 2.}  Calculated dissociation curves for H$_2$ at
the HF (top) and B3LYP (bottom) levels using both a spin and symmetry
restricted approach and a spin and symmetry unrestricted approach.  For
both computational levels the restricted approach is seen to dissociate
to an incorrect higher limit.
\bigskip
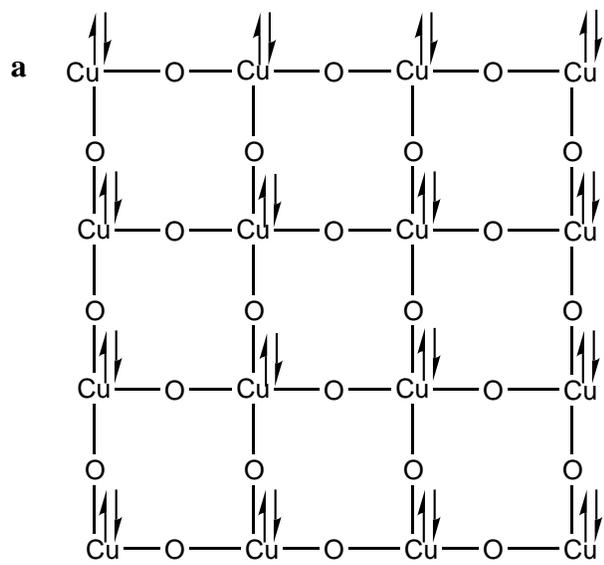
\noindent{\bf Figure 3.}  {\bf a}  Schematic description of Cu spin couplings
under the mean-field approximation.  Each Cu site is 50\% alpha spin and
50\% beta spin.  {\bf b}  Schematic description of the antiferromagnetic
state where alternating Cu sites are either alpha spin or beta spin.  {\bf c}
and {\bf d}  Two schematic descriptions of the paramagnetic state where
a given Cu site may be spin paired with any of the four adjacent Cu sites.
\bigskip
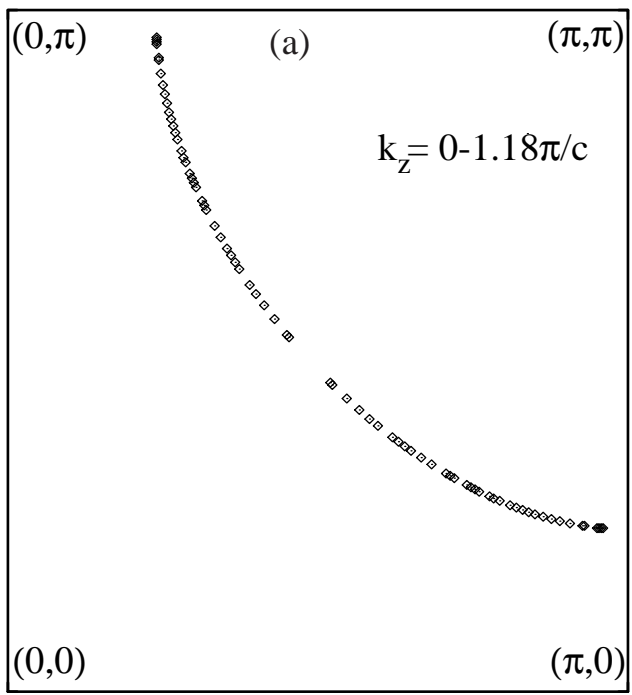
\noindent{\bf Figure 4.} 3-D Fermi surface for optimally doped
La$_{1.85}$Sr$_{0.15}$CuO$_4$.  Cross sections of this Fermi surface are
given at {\bf a} $k_z = 0$, {\bf b} $k_z = 1.3{\pi \over c}$, {\bf c}
$k_z = 1.54{\pi \over c}$, and {\bf d} $k_z = 2{\pi \over c}$.  Electrons
begin to come out of the second band at $k_z = 1.54{\pi \over c}$ allowing
the formation of interband Cooper pairs in the vicinity of the band
crossing.
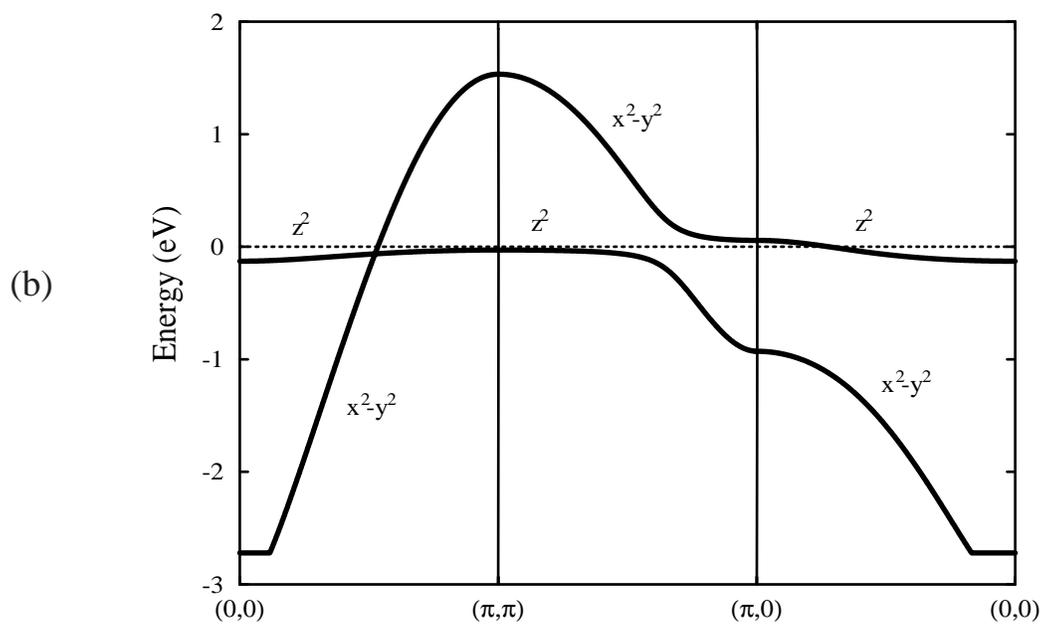
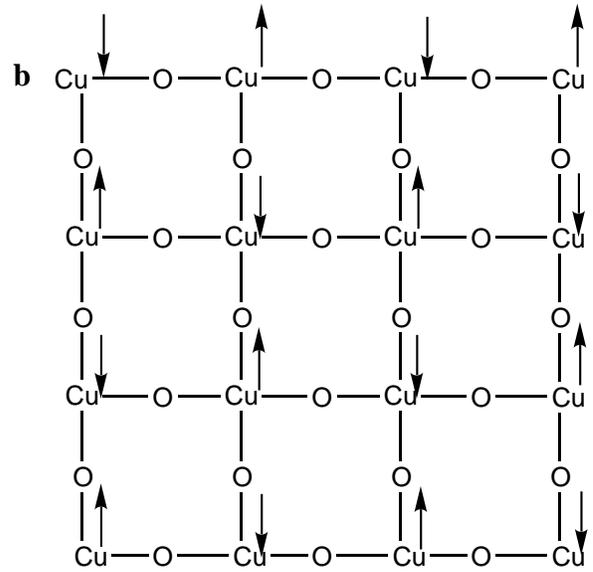
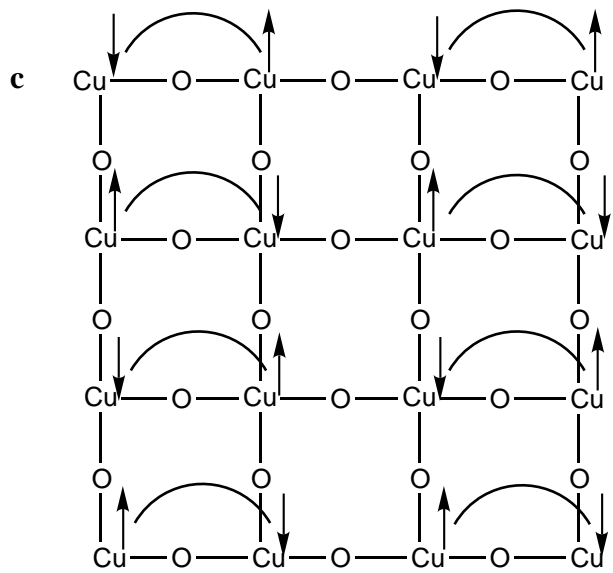
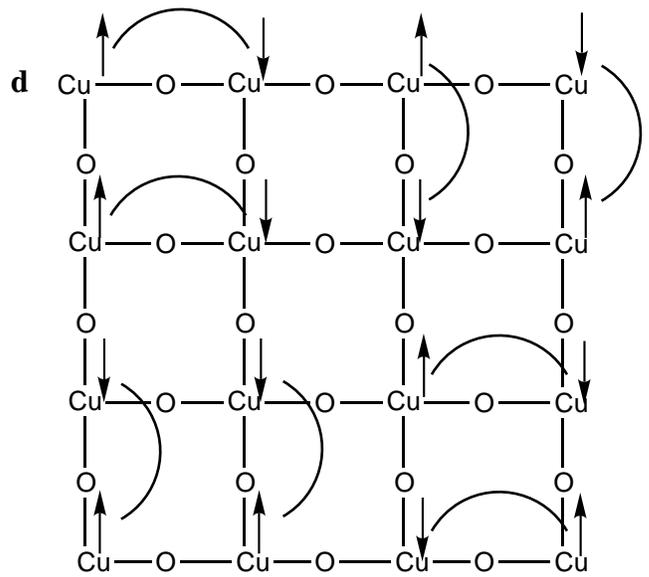
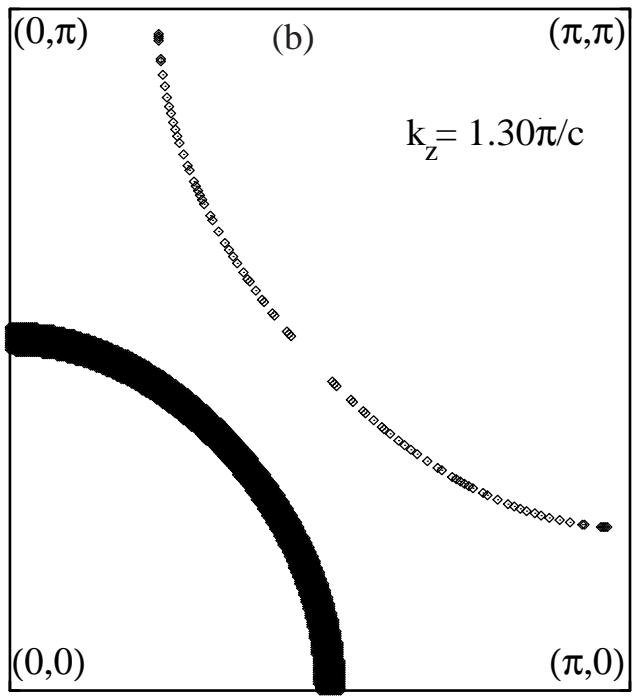
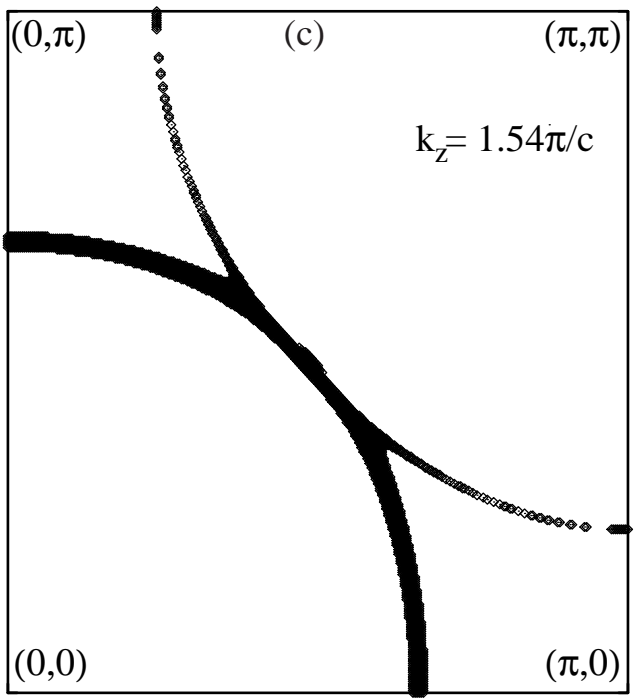
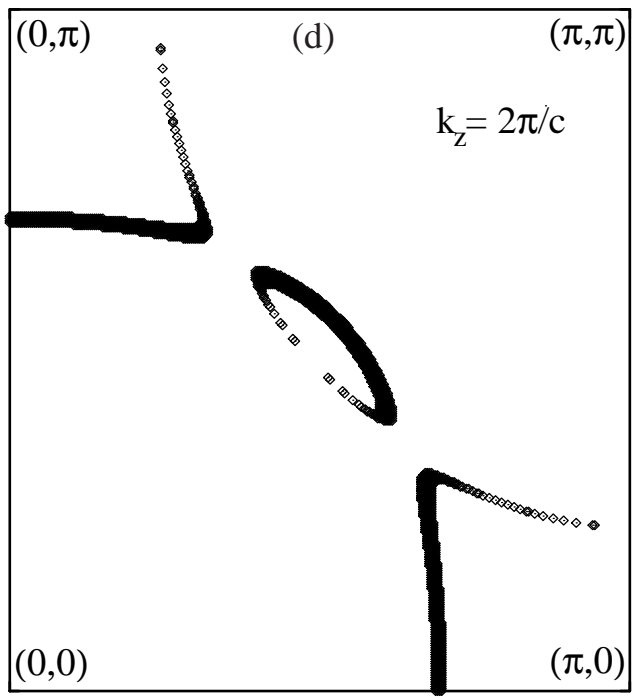
\end